\documentclass[prd,twocolumn,floats,floatfix,nofootinbib,showpacs]{revtex4}
\usepackage{graphicx}
\usepackage{dcolumn}
\usepackage{amssymb}
\usepackage{bm}
\usepackage{amsmath}
\usepackage{multirow}
\usepackage{color}

\begin{document}
\title{Formal analogies between gravitation and electrodynamics}

\author{E. Goulart}
\email{egoulart@cbpf.br}

\author{F. T. Falciano}
\email{ftovar@cbpf.br}
\affiliation{Instituto de Cosmologia Relatividade e Astrofisica ICRA -- CBPF, \\ Rua Xavier Sigaud, 150, Urca, 22290-180, Rio de Janeiro, Brazil}
\date{\today}
\begin{abstract}
We develop a theoretical framework that allows us to compare electromagnetism and gravitation in a fully covariant way. This new scenario does not rely on any kind of approximation nor associate objects with different operational meaning as it's sometime done in the literature. We construct the electromagnetic analogue to the Riemann and Weyl tensors and develop the equations of motion for these objects. In particular, we are able to identify precisely how and in what conditions gravity can be mapped to electrodynamics. As a consequence, many of the gemometrical tools of General Relativity can be applied to Electromagnetism and vice-versa. We hope our results would shed new light in the nature of electromagnetic and gravitational theories.\\
\end{abstract}
\pacs{ }

\maketitle

\section{Introduction}

Right from the beginning, as soon as Einstein proposed his theory, it became an interesting practice to compare the differences and similarities between General Relativity (GR) and Electrodynamics (EM) Ref.'s \cite{Ruffini}-\cite{Pugh}. Nowadays, this is not anymore only a theoretical challenge Ref.\cite{exemplos} but it also attracts much attention of experimental  relativists Ref.\cite{testes}. Notwithstanding very interesting works in the literature, there is still no consensus on how to precisely compare these two theories. 

A priori, a given theory can be formulated in several ways\footnote{ Newtonian Mechanics, for example, has two other formulations, namely, Lagrangian and Hamiltonian formulations.}. Regardless of their equivalence, each formulation has its own peculiarity that makes it more suitable to deal with certain questions. In addition, different  formulations generally increase our understanding of that class of phenomena. The choice of a specific formalism, that could be only a matter of convenience, becomes a crucial point if one wishes to compare different theories. An accurate comparison is only possible if we are able to identify and associate pairs of objects that play similar roles in each theory. This is not an easy task and it is intrinsically related to the framework chosen to work with.


Assuming that Maxwell's Theory and Einstein's General Relativity are the proper theories to describe electromagnetic and gravitational interactions, we can formulate four main challenges that has to be to be addressed if one wishes to correlate them:

\begin{enumerate}
\item[ i)] Electromagnetic fields are described by a vector potential $A^{\mu}$ while gravitation is described by a second rank symmetric tensor $g_{\mu\nu}$;

\item[ ii)] Electromagnetic interactions depend on the charge mass ratio of test particles while the universal character of gravitation allow us to geometrize it;

\item[ iii)] The electric and magnetic fields are defined as first derivatives of the vector potential but it is impossible to construct a tensor with only first derivatives of the metric. This is related to the fact that in GR gravity can be locally cancelled by a coordinate transformation;

\item[ iv)] Electrodynamics solutions satisfy the superposition principle while Einstein's equations are highly non-linear and involved. 
\end{enumerate}

The traditional approaches dealing with these issues can be divided in two main groups depending if they privilege the kinematical or the dynamical features of the theories. The kinematical approach is mainly concerned with the association between equivalent objects (objects that play the same role) while the dynamical one focus on the mathematical structure and the symmetries hidden in the dynamical system of equations. These two approaches can be sketched as follows: 

\textit{Kinematical approach} - this approach is based on the weak field limit of linearized Einstein's equations and deals only with slow moving test particles ($|v^{i}| \ll c$). In this limit, it's possible to choose a specific gauge that puts Einstein's equations in a similar form to Maxwell's equations. In addition, the geodesic equation is ``transformed'' into a newtonian second law with the Christofell's symbol playing the role of a Lorentz force Ref. \cite{Wein-Ohan}. This framework has the advantage to compare only objects with the same physical dimensions. The vector potential $A^{\mu}$ is related to the metric $g_{\mu\nu}$ while the Christofell symbol $\Gamma^{\alpha}_{\mu \nu}$ is associated to the electric $E^{\mu}$ and magnetic $H^{\mu}$ fields. In this scenario, it's possible to apply well known electromagnetic solutions into gravitation with only minor extrapolations Ref. \cite{exemplos}. On the other hand, this method is clearly not covariant and, since it's only valid for a given gauge in the linear limit, it does not properly compare these theories. Furthermore, even ignoring the gauge problem, the approximative nature of this analogy could give the wrong idea that it is only valid in the linear limit.

\textit{Dynamical approach} - there is a formulation of GR, known as Quasi-Maxwellian, that furnishes a dynamic to the Weyl tensor. Lichnerowicz theorem Ref. \cite{Lichnerowicz} guarantees that this system of equations reproduce GR if Einstein's equations are valid in a given hypersurface. In fact, this is a first order system for the irreducible parts of the Weyl tensor Ref.'s \cite{Ellis,Nov2}. The decomposition of the gravitational equations into its four possible projections (so called quasi-maxwellian equations) reproduces in some sense the symmetries of Maxwell's equations for the electric and magnetic part of the Weyl tensor. Contrary to the kinematical approach, this is a fully covariant formalism but, as it will be shown in section \ref{QMpicture}, it fails precisely because it compare completely different objects.

Note that both approaches adapt the gravitational framework keeping the electromagnetism intact. Instead, we propose a new framework that is equivalent to Maxwell's equations but has a mathematical structure similar to the Quasi-Maxwellian formulation. Since we are re-writing electrodynamics in a way similar to GR, we shall call this formalism Quasi-Einsteinnian picture of Maxwell's equations.

This new approach is fully covariant and avoid all the above mentioned problems. In particular, it associates only equivalent objects in the sense that they are of the same physical dimension and operational meaning. In this picture, we can also construct the electromagnetic analogue to the Riemann and Weyl tensors and, most important of all, we can specify precisely when and how electrodynamics differs from gravitational interaction. This control is specially important to identify the situations when we can perfectly map one theory into another.

After specifying some basic mathematical machinery in the next section, we shall briefly describe the main properties of the Quasi-Maxwellian formalism in section \ref{QMpicture}. In section \ref{QEpicture} we construct the Quasi-Einsteinnian formalism and analyse its fundamental aspects. Finally, section \ref{conclusion} is devoted to some final remarks.\\

\section{Mathematical Preliminaries}

The propose of this introductory section is to define some relevant objects and to fix our notation. The space-time signature is chosen to be $(+\, -\, -\, -)$, every time-like congruence $V^{\mu}$ is considered normalised in the sense that $V^{\mu}V_{\mu}=1$ and the symmetric and anti-symmetric parts of a given tensor $A_{\mu\nu}$ are written as $A_{(\mu\nu)}\equiv A_{\mu\nu}+A_{\nu\mu}$ and $A_{[\mu\nu]}\equiv A_{\mu\nu}-A_{\nu\mu}$, respectively. We can define two useful tensors: the Levi-Civita completely anti-symmetric tensor $\eta_{\alpha\beta\mu\nu}$ that satisfy the relations
\begin{eqnarray*}
\eta ^{\mu \nu \varepsilon \lambda}\eta _{\mu \xi \sigma \rho}=-\delta^{\nu \varepsilon \lambda}_{\xi \sigma \rho} \quad ,\\
\eta ^{\mu \nu \varepsilon \lambda}\eta _{\mu \nu \sigma \rho}=-2\delta^{\varepsilon \lambda}_{\sigma \rho} \quad ,
\end{eqnarray*}
and a four rank tensor constructed with the space-time metric and defined by $g_{\alpha\beta\mu\nu}\equiv g_{\alpha\mu}g_{\beta\nu}-g_{\alpha\nu}g_{\beta\mu}$.

As it is well known, the Maxwell equations can be cast in a manifest covariant form by introducing an anti-symmetric tensor $F^{\mu \nu}$ and its dual $F{^{\, \,   \ast}_{\mu \nu}}\equiv \frac{1}{2}\eta _{\mu \nu \varepsilon \lambda}F^{\varepsilon \lambda}$. For a given charge distribution characterised by the four current $J^{\mu}$, the Maxwell's equations are written as
\begin{eqnarray}
&&F_{ \phantom a \phantom a ; \nu}^{\mu \nu}=J^{\mu}\quad ,\label{max1}\\
&&F_{ \phantom a \phantom a ; \nu}^{^{\; \ast}_{\mu \nu}}=0 \quad , \label{max2}
\end{eqnarray}
The kinematical relations given by equation (\ref{max2}) is equivalent to $F^{\left[\alpha \beta ; \lambda \right]}=0$ which allows us to define an electromagnetic potential $A_{\mu}$ such that $F_{\mu \nu}=\partial_{[\mu}A_{\nu]}$. In this context, we understand the electric and magnetic fields as projections of the $F^{\mu \nu}$ tensor along the observer world-line $V^{\nu}$, i.e $E^{\mu}\equiv F^{\mu}_{\phantom a \nu}V^{\nu}$ and $H^{\mu}\equiv \eta^{\mu}_{\phantom a \varepsilon \alpha \beta}V^{\varepsilon}F^{\alpha \beta}$, or inversely we can re-write $F^{\mu \nu}\equiv E^{\left[\mu\right.}V^{\left.\nu\right]}+\frac{1}{2}\eta^{\mu \nu}_{\phantom a \phantom a \alpha \beta}H^{\alpha}V^{\beta}$.

To re-obtain the original, non-covariant, form of Maxwell's equations it suffices to project equations (\ref{max1})-(\ref{max2}) along and perpendicularly to the time-like congruence $V^{\mu}$. This canonical procedure is exactly what we will use to construct the proper framework to establish the analogies between gravitation and electromagnetism.

We finally recall that the Weyl Conformal tensor in an arbitrary 4-dimensional space-time is defined as
\begin{equation}
W^{\alpha\beta}_{\phantom a \phantom a\mu\nu}=R^{\alpha\beta}_{\phantom a \phantom a\mu\nu}+\frac{1}{2} R^{[\alpha}_{\phantom a [\mu}\delta^{\beta]}_{\phantom a\nu]}-\frac{1}{6}R \, g^{\alpha\beta}_{\phantom a\phantom a\mu\nu} \quad ,
\end{equation}
from where we can directly deduce that this tensor is traceless $W^{\alpha}_{\phantom a  \alpha \beta \mu}=W^{\alpha}_{\phantom a\beta\alpha\mu}=W^{\alpha}_{\phantom a\beta \mu \alpha}=0$  and satisfies the symmetry relations $W_{\alpha\beta\mu\nu}=\frac{1}{2}W_{[\mu\nu]\alpha\beta}$ and $W_{\alpha[\beta\mu\nu]}=0$ Ref. \cite{Choquet}.\\

\section{Quasi-Maxwellian picture of Einstein's equations}\label{QMpicture}

In this section we will briefly discuss some properties of the Quasi-Maxwellian formulation of General Relativity. This third order\footnote{The Quasi-Maxwellian equations are third order with respect to the metric tensor but, as will see later, it can also be interpreted as a first order formalism for the irreducible parts of the Weyl tensor $\mathcal{E}_{\alpha\beta}$ and $\mathcal{H}_{\alpha\beta}$.} formalism of Einstein's equations was first suggested by Matte Ref. \cite{Matte} and sequentially developed by Jordan, Ehlers, Kundt, Trumper, Bel, Lichnerowicz and others in the late fifties Ref.'s \cite{Lichnerowicz, Jordan, Kundt}. Our main goal will be to retrace the standard arguments found in the literature and to analyse how the symmetries of the Weyl tensor and its dual share, in a sense, some similitude with the electromagnetic tensor $F_{\mu\nu}$. 

First of all, note that if we take Bianchi identities, which are geometrical properties of any Riemannian space-time, and assume that Einstein's equations are valid, it can be shown that
\begin{equation}\label{weyldiv}
W^{\alpha\beta\mu\nu}_{\phantom a \phantom a \phantom a \phantom a ;\nu}=-\frac{1}{2}T^{\mu [\alpha;\beta]}+\frac{1}{6} g^{\mu[\alpha}T^{,\beta]}\quad ,
\end{equation}
where $T^{\mu \alpha}$ is the energy-momentum tensor describing the matter field and $T$ its trace.

The core of the Quasi-Maxwellian formalism, which was proved by Lichnerowicz, is that if Einstein's equations are valid in a given spatial hypersurface $\Sigma$ than equation (\ref{weyldiv}) propagates it through the whole manifold. In other words, in a Riemannian space-time, if $G_{\mu\nu}=-\kappa \, T_{\mu\nu}$ on a single hypersurface than, necessarily, it also holds everywhere .

This beautiful result allows one to reformulate in a powerful way many applications and issues of general relativity such as perturbation theory applied to cosmology Ref. \cite{Hawk}, emission of gravitational waves Ref. \cite{Bel}, internal symmetries of Einstein's field equations Ref. \cite{Salim}, and the hamiltonian formulation of general relativity Ref. \cite{Novello4}.

The next step in the Quasi-Maxwellian formalism, and precisely when the analogies with electrodynamics comes into play, is to construct all the possible projections of equation (\ref{weyldiv}) with respect to a given observer $V^{\mu}$ and its rest space $h_{\mu \nu}\equiv g_{\mu \nu}-V_{\mu}V_{\nu}$. If we define $\mathcal{E}_{\alpha\mu}\equiv -W_{\alpha\beta\mu\nu}V^{\beta}V^{\nu}$ and $\mathcal{H}_{\alpha\mu}\equiv -W{^{\phantom a \ast}_{\alpha\beta\mu\nu}}V^{\beta}V^{\nu}$ the Weyl conformal tensor admits the following representation
\begin{eqnarray}\label{weylEH}
W_{\alpha\beta\rho\sigma}&=&
(\eta_{\alpha\beta\mu\nu}\eta_{\rho\sigma\lambda\tau}-g_{\alpha\beta\mu\nu}g_{\rho\sigma\lambda\tau})V^{\mu}V^{\lambda}\mathcal{E}^{\nu\tau}+\nonumber \qquad \quad \\
&& + \, (\eta_{\alpha\beta\mu\nu}g_{\rho\sigma\lambda\tau}-g_{\alpha\beta\mu\nu}\eta_{\rho\sigma\lambda\tau})V^{\mu}V^{\lambda}\mathcal{H}^{\nu\tau}\, \, . \qquad
\end{eqnarray}

It's straightforward to show that $\mathcal{E}_{\mu\nu}$ and $\mathcal{H}_{\mu\nu}$ are irreducible, traceless, symmetric and orthogonal with respect to the congruence. These two tensors are known in the literature as the electric ($\mathcal{E}_{\mu\nu}$) and the magnetic ($\mathcal{H}_{\mu\nu}$) parts of the Weyl tensor because they are related to the conformal curvature in a similar way as the electromagnetic fields are related to $F_{\mu\nu}$. The system of equations for the irreducible parts of the Weyl tensor can be found by the four independent equations given by the projections
\begin{eqnarray*}
&&\begin{array}{l}
\mbox{Two vectorial}\\
\mbox{\hspace{.5cm} equations}
\end{array} \left\{
\begin{array}{l}
W^{\phantom a\phantom a \mu \nu}_{\alpha \beta \phantom a \phantom a \;  ;\nu}\, V^{\beta}V_{\mu} \qquad \\
\\
W{^{\; \,  \ast \phantom a \mu \nu}_{\alpha \beta \phantom a \phantom a \;  ;\nu}}\, V^{\beta}V_{\mu}\\
\end{array} \right.\\
&&\\
&&\begin{array}{l}
\mbox{Two tensorial}\\
\mbox{\hspace{.5cm}equations}
\end{array} \left\{
\begin{array}{l}
W{^{\phantom a \phantom a \; \mu \nu}_{\alpha \beta \phantom a \phantom a \;  ;\nu}}\, V^{\beta}h^{\alpha}_{\phantom a (\sigma}h_{\rho) \mu}\\
\\
W{^{\; \,  \ast \phantom a \mu \nu}_{\alpha \beta \phantom a \phantom a \;  ;\nu}}\, V^{\beta}h^{\alpha}_{\phantom a (\sigma}h_{\rho) \mu}\\
\end{array}
\right.\\
\end{eqnarray*}

This system is not only numerically equal to Maxwell's equations but they have a mathematical structure very similar to electrodynamics. These similarities are usually taken as the starting point to construct an analogy between electromagnetism and gravitation. In the next sub-section we will resume the arguments generally given in the literature and argue why we consider this direct analogy between $F^{\mu \nu}$ and $W_{\alpha \mu \beta \nu}$ misleading.\\

\subsection{Misleading analogy between $F^{\mu \nu}$ and $W_{\alpha \mu \beta \nu}$}\label{misleading}

As it was argued above, General Relativity written in the Quasi-Maxwellian form has a mathematical structure that allows one to envisage an analogy between electromagnetism and gravitation. But, to do so, we are forced to associate the electromagnetic tensor $F^{\mu \nu}$ with the Weyl tensor $W_{\alpha \mu \beta \nu}$. Furthermore, the electric $\mathcal{E}_{\mu\nu}$ and the magnetic $\mathcal{H}_{\mu\nu}$ parts of the Weyl tensor have to be associated with the electric $E^{\mu}$ and the magnetic $H^{\mu}$ parts of $F^{\mu \nu}$.

Besides the similarities in the mathematical structure of these two system of equations, there are also two other reasons that seduce us to develop this analogy. If we consider a source-free  region ($T^{\mu \nu}=0$), the Quasi-Maxwellian equations have an additional internal symmetry that plays (as it is generally argued) the same role as the dual symmetry of electrodynamics. As a matter of fact, in vacuum, the Quasi-Maxwellian equations are invariant under an arbitrary field rotation
\begin{eqnarray}\label{rot}
\left(\begin{array}{l}
\widetilde{\mathcal{E}}_{\alpha \beta}\\
\\
\widetilde{\mathcal{H}}_{\alpha \beta}\\
\end{array}\right)&=&\left(\begin{array}{cc}
\cos{\theta}&\sin{\theta}\\
&\\
-\sin{\theta}&\cos{\theta}\\
\end{array}
\right)\left(
\begin{array}{l}
\mathcal{E}_{\alpha \beta}\\
\\
\mathcal{H}_{\alpha \beta}\\
\end{array}\right) \quad .
\end{eqnarray}

In addition, the invariants constructed with $F^{\mu \nu}$ and $W_{\alpha \beta \mu \nu}$, respectively, have the same structure. All these similarities are resumed in the table below \footnote{For some interesting detailed discussion on other properties see Ref. \cite{Nov3}.}.

\begin{displaymath}
\begin{array}{|c|c|c|}
\cline{2-3}
\multicolumn{1}{c|}{}&\multirow{2}{*}{\textbf{\emph{Maxwell}}}&\multirow{2}{*}{\textbf{\emph{Quasi-Maxwellian}}}\\ 
\multicolumn{1}{c|}{}& & \\ 
\hline
\multirow{1}{*}{variables}&F_{\mu\nu}  & W_{\alpha\beta\mu\nu} \\ 
\hline
\multirow{2}{*}{irreducibles}& E^{\alpha} & \mathcal{E^{\alpha\beta}} \\ 
\cline{2-3}
& H^{\alpha} & \mathcal{H^{\alpha\beta}} \\ 
\hline
\multirow{2}{*}{invariants}&E^{\alpha}E_{\alpha}-H^{\alpha}H_{\alpha} × & \mathcal{E_{\alpha\beta}}\mathcal{E^{\alpha\beta}}-\mathcal{H^{\alpha\beta}}\mathcal{H_{\alpha\beta}} \\ 
\cline{2-3}
&E_{\mu}H^{\mu} &2\mathcal{E_{\alpha\beta}}\mathcal{H^{\alpha\beta}} \\
\hline
dual & E_{\alpha}\longrightarrow H_{\alpha}      &  \mathcal{E}_{\alpha\beta}\longrightarrow \mathcal{H}_{\alpha\beta}\\
symmetry $\; $&H_{\alpha}\longrightarrow -E_{\alpha}  & \mathcal{H}_{\alpha\beta}\longrightarrow -\mathcal{E}_{\alpha\beta}\\
\hline
\end{array}
\end{displaymath}\\

Unfortunately, this analogy compares two completely distinct objets. First of all, one can immediately see that $F^{\mu \nu}$ and $W_{\alpha \beta \mu \nu}$ don't have the same dimensional unit. The $F^{\mu\nu}$ tensor plays the role of a force for charged particles while $W_{\alpha \beta \mu \nu}$ measures the relative movement of two particles in vacuum. Note also that Maxwell's equations are second order equations for the electromagnetic potential while the Quasi-Maxwellian equations are third order for $g_{\mu\nu}$. In addition, it is not clear the meaning of the gravitational dual symmetry and its relation to the electromagnetic case. Actually, it is natural to suppose that we must correlate only objects that are of the same order with respect to the electromagnetic potential and the metric tensor. It is clear that this is not the case if we insist to associate $F^{\mu \nu}$ with $W_{\alpha \beta \mu \nu}$. Another strong argument against this comparison comes from the geodesic deviation equation. In vacuum, the equation for the connecting vector $\eta^{\alpha}$ is
\begin{equation}\label{gde1}
\dfrac{D^{2}\eta^{\alpha}}{Ds^{2}}=W^{\alpha}_{\phantom a \beta \mu \nu}\eta^{\mu}V^{\beta}V^{\nu}=\cal{E}^{\alpha}_{\phantom a \, \mu}\, \eta^{\mu} \quad .
\end{equation}

As it is shown below, eq. (\ref{ade1}), the equation that describe the vector connecting two initially parallel charged particles is analogous to equation (\ref{gde1}) but replacing the Weyl tensor $W^{\alpha}_{\phantom a \beta \mu \nu}$ by $F^{\alpha}_{\phantom a \, \beta \, ; \mu}$. Note that, these tensors are both of the same order in their potentials. $W^{\alpha}_{\phantom a \beta \mu \nu}$ have third order derivative of the metric $g_{\mu \nu}$ just as $F^{\alpha}_{\phantom a \, \beta \, ; \mu}$ have third order derivative of the electromagnetic potential $A^{\mu}$.\\

\section{Quasi-Einsteinian picture of Maxwell's equations}\label{QEpicture}

If we want to construct a common framework to establish a dialogue between two different theories, such as general relativity and electromagnetism, first of all, it is mandatory to specify a rule to associate their objects. We understand that this rule should be defined not just by a pure analogy of similar objets, $F_{\mu\nu}$ and $W_{\alpha\beta\mu\nu}$ for instance, but rather by their physical-mathematical operational meaning.

One could propose to associate General Relativity and Electromagnetism by comparing the test particle's equation of motion in the presence of a gravitational field (geodesic equation) with the test charged particle's equation of motion in an electromagnetic field (Lorentz force).
\begin{eqnarray*}
\mbox{geodesic eq.}&:&\frac{\mathrm{d}^{2}x^{\alpha}}{\mathrm{d}s^{2}}=\Gamma^{\alpha}_{\phantom a \mu \nu} \, V^{\mu}V^{\mu}\\
\mbox{Lorentz force}&:&\frac{\mathrm{d}^{2}x^{\alpha}}{\mathrm{d}s^{2}}=F^{\alpha}_{\phantom a \, \mu} \,V^{\mu}\\
\end{eqnarray*}

But then, we would be forced to relate $F^{\mu\nu}$ to $\Gamma^{\alpha}_{\phantom a\mu\nu}$. It is reasonable to pursue this association only in the weak field limit of General Relativity where it is possible to arrange the first order equations exactly as Maxwell's equations for a stationary electric and magnetic fields Ref.'s \cite{exemplos, testes}. This analogy can be useful to study gravimagnetic fields or frame dragging phenomena Ref. \cite{Gravitodynam}. Notwithstanding its utility in some special cases, the Christoffel's symbol is not a tensor. Any kind of comparison made in this way shall be \emph{fortuitous and gauge dependent}. In effect, General Relativity does not admit a tensor only with first derivatives of the metric tensor $g_{\mu\nu}$. Thus, it is impossible to construct, in a covariant way, a framework to compare General Relativity and Electromagnetism using first order derivatives of the potentials $A^{\mu}$ and $g_{\mu \nu}$.

Up to this point, we have only briefly resumed previous works and contented ourselves to argue against their inappropriateness. Now, we propose a different way to look into Maxwell's equations which will deal with third order derivatives of the electromagnetic potential $A^{\mu}$. This formalism contains a very natural generalisation of the Riemann and Weyl tensors for electrodynamics and it provides the necessary framework to compare electrodynamics and General Relativity on equal footing. In what follows we will assume that the charge mass ratio is constant $\frac{q}{m}=const.$. Instead of looking to the geodesic equation, we focus in the geodesic deviation equation and the equation for the deviation vector $\eta^{\mu}$ connecting the world-lines of two charged particles with parallel initial tangent vectors ($V^{\mu}_{\phantom a \, ;\nu}\, \eta^{\nu}=0$)
\begin{equation}\label{ade1}
\dfrac{D^{2}\eta^{\alpha}}{Ds^{2}}=F^{\alpha}_{\phantom a\mu;\nu}V^{\mu}\eta^{\nu} \quad .
\end{equation}

The above equation shows us that the natural choice is to identify $F^{\alpha}_{\phantom a\mu;\nu}$ as the electromagnetic analogue of the Riemann tensor. Once the tensor catalogue we will construct has a direct counterpart in General Relativity, even-though, in principle, they have nothing to do with geometry, we suggest to call this framework Quasi-Einsteinian picture of Electrodynamics.\\

\subsection{Some useful decompositions and projections}

In general, $F_{\alpha \beta \, ;\lambda}$ has a non-zero trace, see eq.(\ref{max1}). Accordingly to General Relativity, the electromagnetic analogue of the Weyl tensor should be defined as the traceless part of the electromagnetic analogue to the Curvature tensor. Thus, we define 

\begin{equation}\label{defI}
I_{\alpha \beta \lambda} \equiv F_{\alpha \beta ;\lambda}-\frac{1}{3}\left(F_{\alpha}\, g_{\beta \lambda}-F_{\beta}\, g_{\alpha \lambda} \right) \quad ,
\end{equation}
where $F_{\alpha}\equiv F_{\alpha \phantom a ; \varepsilon}^{\phantom a \varepsilon}$. Note that, $I_{\alpha \beta \lambda}$ has inherited the anti-symmetry in the first two index $I_{\alpha\beta\mu}=\frac{1}{2}I_{[\alpha\beta]\mu}$. Using Maxwell's equations, it is also straightforward to show that $I_{\left[\alpha \beta \lambda\right]}=0$. By construction, $I_{\alpha \beta \lambda}$ does not depend on the current field $J^{\mu}$. Nonetheless, there are integrability conditions relating $I_{\alpha \beta \lambda}$ to derivatives of the current vector $J_{\mu \, ; \alpha}$ in the same way as $W_{\alpha \mu \beta \nu}$ is connected to derivatives of the energy-momentum tensor by the Bianchi identities.

With the help of a time-like congruence $V^{\mu}$ and its associated projector $h_{\mu \nu}\equiv g_{\mu \nu}-V_{\mu}V_{\nu}$, we can decompose the $I_{\alpha \beta \lambda}$ and its dual tensor as

\begin{eqnarray}
&&I_{\alpha\beta\mu}=-V_{[\alpha}Q_{\beta]\mu}-\eta_{\alpha\beta\lambda\varepsilon}Z^{\lambda}_{\phantom a \mu}V^{\varepsilon} \quad ,\label{IQZ}\\
&&I{^{\phantom a \ast}_{\alpha \beta \mu}}=-V_{[\alpha}Z_{\beta]\mu}+\eta_{\alpha\beta\lambda\varepsilon}Q^{\lambda}_{\phantom a \mu}V^{\varepsilon} \quad ,\label{IZQ}
\end{eqnarray}
where $Q_{\alpha \beta} \equiv I_{\alpha \varepsilon \beta}\, V^{\varepsilon}$ and $Z_{\alpha \beta} \equiv I{^{\, \,  \ast}_{\alpha \varepsilon \beta}}V^{\varepsilon}$. These are traceless tensors ($Q^{\beta}_{\phantom a \beta}=Z^{\beta}_{\phantom a \beta}=0$) with the first index projected on the hypersurface, i.e. $Q_{\alpha\mu}V^{\alpha}=Z_{\alpha\mu}V^{\alpha}=0$. It's interesting to notice, as can be seen from equations (\ref{IQZ})-(\ref{IZQ}), that this system satisfy the same kind of rotational symmetry as the Quasi-Maxwellian one. The system is invariant with respect to a rotation in the $Q_{\alpha \beta}$ x $Z_{\alpha \beta}$ plane analogous to equation eq. (\ref{rot}). In addition, the projection of only one index of the Weyl tensor along the time-like congruence can be written in a very similar way to equation (\ref{IQZ})

\begin{equation}\label{WV}
W_{\alpha \beta \mu \nu}V^{\nu}=V_{[\alpha}\mathcal{E}_{\beta]\mu}+\eta_{\alpha\beta\lambda\varepsilon}\mathcal{H}^{\lambda}_{\phantom a \mu}V^{\varepsilon} \quad .\\
\end{equation}

Nevertheless, there is a crucial difference that forbid us to relate them. The $Q_{\alpha \beta}$ and $Z_{\alpha \beta}$ are not irreducible parts of $I_{\alpha \beta \mu}$ such as $\mathcal{E}_{\alpha \beta}$ and $\mathcal{H}_{\alpha \beta}$ are of $W_{\alpha \beta \mu \nu}$. The proper analogy between Electromagnetism and General Relativity is made using the irreducible parts of $I_{\alpha \beta \mu}$. As we shall see, tidal effects are considerably more sophisticated in electrodynamics than in gravitation. This is a remarkable result related to the fact that electrodynamics is a vectorial theory while General Relativity is a pure tensorial theory. If this is the case, it's expected that alternative gravitational theories such as TeVeS Ref. \cite{Teves} should also produce some of these electromagnetic interesting features with respect to tidal forces.\\

\subsection{The irreducible parts of the electromagnetic analogue Weyl tensor}

As equation (\ref{IQZ}) suggests, the $I_{\alpha \beta \mu}$ tensor can be completely described by the irreducible parts of $Q_{\alpha\beta}$ and $Z_{\alpha\beta}$.\footnote{Actually, there is also the possibility to project all three index of $I_{\alpha \beta \mu}$ on the hypersurface but it can be shown that this new tensor is completely determined by the four tensors that compose the irreducible parts of $Q_{\alpha\beta}$ and $Z_{\alpha\beta}$.} Once they are not completely on the hypersurface orthogonal to $V^{\mu}$ we have

\begin{eqnarray}
&& Q_{\alpha\beta}=\widehat{Q}_{\alpha \beta}+Q_{\alpha}V_{\beta}\\
&& Z_{\alpha\beta}=\widehat{Z}_{\alpha \beta}+Z_{\alpha}V_{\beta}
\end{eqnarray}
where $\widehat{Q}_{\alpha \beta} \equiv Q_{\alpha \varepsilon} \, h^{\varepsilon}_{\phantom a \beta}$,
$\widehat{Z}_{\alpha \beta}\equiv Z_{\alpha \varepsilon}\, h^{\varepsilon}_{\phantom a \beta}$, $Q_{\alpha} \equiv  I_{\alpha \varepsilon \rho}\, V^{\varepsilon}\, V^{\rho}$ and $Z_{\alpha} \equiv  I{^{\; \, \ast}_{\alpha \varepsilon \rho}}\, V^{\varepsilon}$. In these equations the $\widehat{\phantom a }$ means that the tensor is entirely restricted to the space-like hypersurface. It is immediate to see that the vector $Q^{\alpha}$ and $Z^{\alpha}$ are also on the hypersurface. The $\widehat{Q}_{\alpha \beta}$ and $\widehat{Z}_{\alpha \beta}$ tensors can still be decomposed relatively to its symmetric and anti-symmetric parts.
\begin{eqnarray*}
\widehat{Q}_{\alpha \beta}&=&E_{\alpha \beta}+\frac{1}{2} \widehat{Q}_{\left[\alpha \beta \right]} \quad , \qquad E_{\alpha \beta}\equiv \frac{1}{2}\widehat{Q}_{\left( \alpha \beta\right)}\quad , \\
\widehat{Z}_{\alpha \beta}&=&H_{\alpha \beta}+\frac{1}{2} \widehat{Z}_{\left[\alpha \beta \right]}\quad , \qquad H_{\alpha \beta}\equiv \frac{1}{2}\widehat{Z}_{\left( \alpha \beta\right)}\quad .
\end{eqnarray*}

We note that the symmetric tensors $E_{\alpha\beta}$ and $H_{\alpha\beta}$ satisfies exactly the same properties of the electric and magnetic parts of Weyl tensor i.e. $E_{\alpha\beta}V^{\beta}=H_{\alpha\beta}V^{\beta}$=0 and $E^{\alpha}_{\phantom a \alpha}=H^{\alpha}_{\phantom a \alpha}=0$. Using their own definitions, the anti-symmetric parts of $\widehat{Q}_{\alpha \beta}$ and $\widehat{Z}_{\alpha \beta}$ can be written in terms of the vectors  $Q^{\alpha}$ and $Z^{\alpha}$ as

\begin{eqnarray*}
\widehat{Q}_{\left[\alpha \beta \right]}=-\eta_{\alpha \beta \varepsilon \lambda}Z^{\varepsilon}V^{\lambda} &\quad \mbox{or} \quad& Z_{\alpha}=\frac{1}{2}\eta_{\alpha \mu \nu \lambda}\widehat{Q}^{\left[\mu \nu \right]}V^{\lambda}\quad , \qquad \\
 \quad \widehat{Z}_{\left[\alpha \beta \right]}=-\eta_{\alpha \beta \varepsilon \lambda}Q^{\varepsilon}V^{\lambda}& \quad \mbox{or} \quad& Q_{\alpha}=\frac{1}{2}\eta_{\alpha \mu \nu \lambda}\widehat{Z}^{\left[\mu \nu \right]}V^{\lambda}\quad . \qquad 
\end{eqnarray*}

Consequently all the information contained in $I_{\alpha \beta \lambda}$ is coded in four tensor that are restricted to the hypersurface. In view of its irreducible parts, $I_{\alpha \beta \mu}$ and its dual are given as
 \begin{eqnarray}
I_{\alpha \beta \lambda}&=&-V_{\left[\alpha\right.}E_{\left.\beta\right]\lambda}+\eta_{\alpha \beta \mu \nu}V^{\mu}H^{\nu}_{\phantom a \lambda}+\frac{1}{2}Q_{\left[\alpha\right.}h_{\left.\beta\right]\lambda}+ \qquad\\
&&+Q_{\left[\alpha\right.}V_{\left.\beta\right]}V_{\lambda}-\eta_{\alpha \beta \mu \nu}Z^{\mu}V^{\nu}V_{\lambda}+\frac{1}{2}V_{ [\alpha} \eta_{\beta] \lambda \mu \nu}Z^{\mu}V^{\nu} \, , \nonumber\\
&&\nonumber\\
I{^{\phantom a \ast}_{\alpha \beta \lambda}}&=&-V_{\left[\alpha\right.}H_{\left.\beta\right]\lambda}-\eta_{\alpha \beta \mu \nu}V^{\mu}E^{\nu}_{\phantom a \lambda}-\frac{1}{2}Z_{\left[\alpha\right.}h_{\left.\beta\right]\lambda}+ \quad\\
&&+Z_{\left[\alpha\right.}V_{\left.\beta\right]}V_{\lambda}+\eta_{\alpha \beta \mu \nu}Q^{\mu}V^{\nu}V_{\lambda}+\frac{1}{2}V_{ [\alpha} \eta_{\beta] \lambda \mu \nu}Q^{\mu}V^{\nu} \, . \nonumber
\end{eqnarray}

The consistency check is made by counting the degrees of freedom for each one of the tensors. Since $I_{\alpha \beta \lambda}$ and its dual are traceless and they are anti-symmetric in the first two indexes, $I_{\alpha \beta \lambda}$ has 16 independent components (we recall that the Weyl tensor in four dimensions has 10 independent components). On the other hand, $E_{\alpha \beta}$ and $H_{\alpha \beta}$ are completely projected and are also traceless ($E_{\phantom a \alpha}^{\alpha}=H_{\phantom a \alpha}^{\alpha}=0$ , $E_{\alpha \beta}V^{\beta}=H_{\alpha \beta}V^{\beta}=0$) so each one of them has 5 independent components. The other 6 missing degrees of freedom are divided between the two vector since they are on the hypersurface ($Q_{\alpha}V^{\alpha}=Z_{\alpha}V^{\alpha}=0$).

Note that vanishing of $I_{\alpha \beta \mu}$ implies the simultaneous vanishing of $E_{\alpha \beta}$, $H_{\alpha \beta}$, $Q_{\alpha}$ and $Z_{\alpha}$. This interesting situation, which it is obviously possible in electrodynamics without the vanishing of the fields $E^{\mu}$ and $H^{\mu}$ themselves, could be understood as the electromagnetic analogue of a conformally flat solution of Einstein's equations.

The distinct character of electrodynamics appears with the two vectors $Q^{\alpha}$ and $Z^{\alpha}$. In fact, the $E_{\alpha \beta}$ and $H_{\alpha \beta}$ are perfectly mapped to $\mathcal{E}_{\alpha \beta}$ and  $\mathcal{H}_{\alpha \beta}$ respectively. The two simplest invariant constructed with $I_{\alpha \beta \mu}$ and its dual are\footnote{Since $I_{\alpha\beta\mu}$ is a traceless tensor only with three indexes, it is impossible to construct any third order invariant.}

\begin{eqnarray*}
I_{\alpha \beta \mu}I^{\alpha \beta \mu}&=&2\left(E_{\mu \nu}E^{\mu \nu}-H_{\mu \nu}H^{\mu \nu}\right)+3\left(Q_{\alpha}Q^{\alpha}-H_{\alpha}H^{\alpha}\right) \qquad \\
I{^{\phantom a \ast}_{\alpha \beta \mu}}I^{\alpha \beta \mu}&=&4E_{\mu \nu}H^{\mu \nu}+2Q_{\alpha}Z^{\alpha} \quad .
\end{eqnarray*}

These expressions are very similar to that presented in the Quasi-Maxwellian picture section except by the vector terms. Note also that there is no cross term between vectors and tensors. We conjecture that a detailed study of all possible independent algebraic invariants might provide a possible classification of electrodynamics configurations in a similar way as Debever used to classify General Relativity Ref. \cite{Deb}.

In terms of these objects, the deviation equation, eq.(\ref{ade1}), in a source-free region assumes the form ($V^{\mu}\eta_{\mu}=0$)

\[
\dfrac{D^{2}\eta^{\alpha}}{Ds^{2}}=\left(E^{\alpha}_{\phantom a \nu}-\frac{1}{2}\eta^{\alpha}_{\phantom a \nu\lambda\rho}Z^{\lambda}V^{\rho}\right)\eta^{\nu} \quad ,
\]
which is exactly eq.(\ref{gde1}) except for the extra vectorial term\footnote{As first shown by Papapetrou Ref.'s \cite{Papapetrou1, Papapetrou2}, the magnetic part of the Weyl tensor $\cal{H}_{\mu \nu}$ influence the trajectory of any particle having angular momentum or spin. In the electromagnetic case there is a complete analogous equation coupling the angular momentum or the spin to $H_{\mu \nu}$ and $Q^{\alpha}$. Thus, the papapetrou equation also strength our attempt to associate the tensor $I_{\alpha\beta\mu}$ to the Weyl conformal tensor.}. We immediately see that, since the tensor $E_{\alpha\beta}$ has exactly the same properties of the electric part of Weyl tensor $\mathcal E_{\alpha\beta}$, electrodynamics will mimic gravitational tidal forces if and only if the vector $Z^{\alpha}$ is zero. 

In the next section we will develop the dynamical system for these irreducible tensors but it's already possible to notice in which cases there is a full analogue between electrodynamics and gravitation.

The covariant formalism constructed here shows that discrepancy between electrodynamics and gravitation comes from the two vector $Q^{\alpha}$ and $Z^{\alpha}$. \emph{If and only if these two vectors vanish simultaneously it's possible to map one theory into another}. We also like to reinforce that this is a complete covariant statement that does not depend in any kind of approximation and compares ``equivalent objects'' from each theory.\\


\subsection{Dynamical equations}

In this section we derive the dynamical equations for the Quasi-Einsteinnian formalism. The full general equations are quite extensive so, for the sake of clarity, here we will assume only inertial observers ($V_{\alpha ;\beta}=0$) in flat space-time ($R_{\alpha \mu \beta \nu}=0$). The generalisation to non-inertial observers in curved space-time is given in appendix \ref{appendixA}.

Similarly to the Quasi-Maxwellian equations, the dynamics for the given system are determined by the divergence of the $I_{\alpha \beta \lambda}$ and its dual tensor. The divergence for $I_{\alpha \beta \mu}$ can be calculated directly. The other dynamical equation can be found by the following relation
\[
\nabla_{\left[ \alpha\right.}I_{\left. \mu \nu \right] \beta}= -\frac{1}{3} \left(\nabla_{\left[ \alpha \right.}F_{\left. \mu \right]}g_{\nu \beta}+\nabla_{\left[ \nu \right.}F_{\left. \alpha \right]}g_{\mu \beta}+\nabla_{\left[ \mu \right.}F_{\left. \nu \right]}g_{\alpha \beta}\right)\quad . \qquad 
\]
It's straightforward to show that
\begin{eqnarray}\label{din1}
I^{\mu \beta \alpha}_{\phantom a \phantom a \phantom a \, ;\beta}&=&\frac{2}{3}F^{\mu ; \alpha} \quad , \quad\\
I_{\phantom a \phantom a \phantom a ;\beta}^{^{\; \, \ast}_{\mu \beta \alpha}}&=&\frac{1}{3}F^{^{\; \ast}_{\mu ; \alpha}}\quad.\label{din2}
\end{eqnarray}

To study the dynamics through the irreducible representations $E^{\mu \nu}$, $H^{\mu \nu}$, $Q^{\alpha}$ and $Z^{\alpha}$ we have to perform the four possible projection along and perpendicular to the congruence defined by $V^{\mu}$. The time derivative of any given tensor is defined as $\dot{\xi}^{\mu}\equiv \xi^{\mu}_{\phantom a ; \alpha}V^{\alpha}$ while the projected four current is written as $j^{\mu}\equiv J^{\alpha}h^{\phantom a \mu}_{\alpha}$. Equation (\ref{din1}) has the following projections

\begin{eqnarray*}
&\mbox{i)}I_{\phantom a \phantom a \phantom a ; \, \beta}^{\mu \beta \alpha}V_{\mu}V_{\alpha}:&-Q^{\beta}_{\phantom a \, ; \beta}=\frac{2}{3}\dot{\rho} \quad \\
&\mbox{ii)}I_{\phantom a \phantom a \phantom a; \beta}^{\mu \beta \alpha}V_{\alpha}h^{\rho}_{\phantom a \mu}:&\dot{Q}^{\rho}-\eta^{\rho \beta}_{\phantom a \phantom a \varepsilon \sigma}Z^{\varepsilon}_{\phantom a  ; \beta}V^{\sigma}=\frac{2}{3}(j^{\rho}\dot{)} \quad\\
&\mbox{iii)}I_{\phantom a \phantom a \phantom a; \beta}^{\mu \beta \alpha}V_{\mu}h^{\lambda}_{\phantom a \alpha}:&-E^{\lambda \beta}_{\phantom a \; ; \beta}-\frac{1}{2}\eta^{\lambda \beta}_{\phantom a \phantom a \varepsilon \sigma}Z^{\varepsilon}_{\phantom a  ; \beta}V^{\sigma}=\frac{2}{3}\rho_{; \alpha}h^{\lambda \alpha}\qquad \\
&\mbox{iv)}I_{\phantom a \phantom a \phantom a; \beta}^{\mu \beta \alpha}h^{\rho}_{\phantom a \mu}h^{\lambda}_{\phantom a \alpha}:&\dot{E}^{\rho \lambda}-\frac{1}{2}\eta^{\rho \lambda}_{\phantom a \phantom a \varepsilon \sigma} \dot{Z}^{\varepsilon}V^{\sigma}-\eta^{\rho \beta}_{\phantom a \phantom a  \varepsilon \sigma}Z^{\varepsilon \lambda}_{\phantom a \phantom a ; \beta}V^{\sigma}=\quad \nonumber\\
&&=\frac{2}{3}\left[\dot{\rho}V^{\lambda}V^{\rho}-\dot{J}^{\rho}V^{\lambda} -\rho^{\, ; \lambda}V^{\rho}+J^{\rho;\lambda}\right] \qquad
\end{eqnarray*}

while equation (\ref{din2}) give us

\begin{eqnarray*}
&\mbox{v)}I_{\phantom a \phantom a \phantom a ;\beta}^{^{\phantom a \ast}_{\mu \beta \alpha }}V_{\mu}V_{\alpha}:&-Z^{\beta}_{\phantom a ; \beta}=0 \quad ,\\
&\mbox{vi)}I_{\phantom a \phantom a \phantom a \, ;\beta}^{^{\phantom a \ast}_{\mu \beta \alpha}}V_{\alpha}h^{\rho}_{\phantom a \mu}:&
\dot{Z}^{\rho}+\eta^{\rho \beta}_{ \phantom a \phantom a  \varepsilon \sigma}Q^{\varepsilon}_{\phantom a \, ; \beta}V^{\sigma}=\frac{1}{3}\eta^{\rho}_{\phantom a \varepsilon \sigma \alpha}J^{\varepsilon \, ; \sigma}V^{\alpha} \; ,\\
&\mbox{vii)}I_{\phantom a \phantom a \phantom a ;\beta}^{^{\phantom a \ast}_{ \mu \beta \alpha}}V_{\mu}h^{\lambda}_{\phantom a \alpha}:&
-H^{\lambda \beta}_{\phantom a \phantom a ;\beta}-\frac{1}{2}\eta^{\lambda \beta}_{\phantom a \phantom a \varepsilon \sigma}Q^{\varepsilon}_{\phantom a \, ; \beta}V^{\sigma}=\\
&&=-\frac{1}{3}\eta^{\lambda}_{\phantom a \varepsilon \sigma \alpha}J^{\varepsilon \, ; \sigma}V^{\alpha} \quad ,\\
&\mbox{viii)}I_{\phantom a \phantom a \phantom a ;\beta}^{^{\phantom a \ast}_{\mu \beta \alpha}}h^{\rho}_{\phantom a \mu}h^{\lambda}_{ \phantom a \alpha}:&\dot{H}^{\rho \lambda } -\frac{1}{2}\eta^{ \rho \lambda}_{\phantom a \phantom a \varepsilon \sigma} \dot{Q}^{\varepsilon}V^{\sigma}+\eta^{\rho \beta}_{\phantom a \phantom a  \varepsilon \sigma}Q^{\varepsilon \lambda}_{\phantom a \phantom a ; \beta}V^{\sigma}=\qquad \\
&&=\frac{1}{3}\left(\eta^{\rho \lambda}_{\phantom a \phantom a \varepsilon \sigma}J^{\varepsilon \, ; \sigma}+V^{\left[ \rho \right.}\eta^{\left. \lambda \right]}_{\phantom a \varepsilon \sigma \alpha}J^{\varepsilon \, ; \sigma}V^{\alpha} \right) \qquad
\end{eqnarray*}

This system is composed of two scalar equations, i) and v), two vectorial equations, ii) and vi), and four tensorial equations, iii), iv), vii) and viii). The four tensorial equations gives the dynamics of the two tensor $E_{\mu \alpha}$ and $H_{\mu \alpha}$ and as expected are completely analogous to the gravitational system except for the terms involving the vectors $Q^{\alpha}$ and $Z^{\alpha}$.

It's also interesting to notice that the remaining four equations defines a closed system for the two vectors. In a source-free region these four equations have exactly the same form as Maxwell's equations but instead of being for the electric and magnetic field itself are for the $Q^{\alpha}$ and $Z^{\alpha}$ vectors. 

Once these system is completely characterised, the next step is to analyse particular solutions. With this formulation of Maxwell's equations we can turn on and off the terms that distinguish gravitation and electrodynamics. In addition, it seems reasonable to admit that many of the technical resources for solving some class of solutions in one of the theories can be used to solve/study the similar type of solution in the other theory. Some very interesting cases will be analysed elsewhere Ref. \cite{falcianogoulart}.\\

\section{Conclusion}\label{conclusion}

In this work, we have constructed a framework where it is possible to properly compare electromagnetism with gravitation. We have chosen to reformulate electrodynamics so that its mathematical structure appears similar to the Quasi-Maxwellian formulation of General Relativity. The main point is that with our formulation the analogies created between these theories compare objects that play the same role in each theory.

For the sake of conciseness, we have only briefly resumed the common approaches already well known in the literature connecting Maxwell's theory to General Relativity. We hope to have convinced the reader that up to now there was no satisfactory framework that incorporate completely both theories.

At this moment, it is imperative to mention the work of L. Filipe Costa and Carlos A. R. Herdeiro Ref. \cite{costaherdeiro}. Our formalism have been developed independently and only recently we were aware of their work. Hence, we decided to maintain the structure of our paper without adding an analysis of their formalism. Actually, their proposal is very similar to ours but there is a crucial difference: they form an analogy between $Q_{\mu \alpha}$ with $\mathcal{E}_{\mu \alpha}$ and  $Z_{\mu \alpha}$ with $\mathcal{H}_{\mu \alpha}$. As we have explained earlier, the proper analogies are made with the irreducible parts of these tensor. Without the irreducible parts of $I_{\alpha \beta \mu}$ it is impossible to discriminate exactly where and how electrodynamics differs from gravitation.

It is worth to resume our formalism in a table similar to the one presented in section \ref{misleading}:
\begin{displaymath}
\begin{array}{|c|c|c|}
\cline{2-3}
\multicolumn{1}{c|}{}&\multirow{2}{*}{\textbf{\emph{Quasi-Einsteinnian}}}&\multirow{2}{*}{\textbf{\emph{Quasi-Maxwellian}}}\\ 
\multicolumn{1}{c|}{}& & \\ 
\hline
variables&I_{\alpha \beta \mu}  & W_{\alpha\beta\mu\nu} \\ 
\hline
\multirow{4}{*}{irreducibles}&Q^{\alpha}& \mbox{\emph{No Analogue}}\\
\cline{2-3}
&Z^{\alpha}&\mbox{\emph{No Analogue}}\\
\cline{2-3}
& E^{\alpha \beta} & \mathcal{E^{\alpha\beta}} \\ 
\cline{2-3}
& H^{\alpha \beta} & \mathcal{H^{\alpha\beta}} \\ 
\hline
\multirow{3}{*}{invariants}&2\left(E_{\mu \nu}E^{\mu \nu}-H_{\mu \nu}H^{\mu \nu}\right)&\multirow{2}{*}{$\mathcal{E_{\alpha\beta}}\mathcal{E^{\alpha\beta}}-\mathcal{H^{\alpha\beta}}\mathcal{H_{\alpha\beta}}$}\\ 
&+\, 3\left(Q_{\alpha}Q^{\alpha}-H_{\alpha}H^{\alpha}\right) &\\
\cline{2-3}
&{4\, E_{\mu \nu}H^{\mu \nu}+2\, Q_{\alpha}Z^{\alpha}} &2\mathcal{E_{\alpha\beta}}\mathcal{H^{\alpha\beta}} \\
\hline
dual & \mbox{symmetry restored }   &  \mathcal{E}_{\alpha\beta}\longrightarrow \mathcal{H}_{\alpha\beta}\\
symmetry& \mbox{only if  $Q^{\mu}=Z^{\mu}=0$}& \mathcal{H}_{\alpha\beta}\longrightarrow -\mathcal{E}_{\alpha\beta}\\
\hline
\end{array}
\end{displaymath}\\

Since we understand that we were developing a new approach to compare Electrodynamics and General Relativity, it was wise to restrict our discussion and only define the main objects and stress the internal structure of the Quasi-Einsteinnian formalism. However, there is still a list of promising routes to be analysed.

We have already mentioned that we expect to learn from all the technical tools developed for one of the theories and to be able to apply into the other. Thus, a formal map between Quasi-Maxwellian to Quasi-Einsteinnian formalism could allow us to gain a new understanding of old solutions. Even more interesting, there is the possibility to construct new classes of solutions, for example, we could look for the electromagnetic analogue of all cosmological solutions such as Friedmann -Robertson -Walker, de Sitter, Kerr or even G\"odel's metric.

We also conjecture that it should be possible to classify the electromagnetic solution in terms of the properties of its invariants. It would be very interesting to compare this classification to the well known Petrov's classification for the Weyl tensor.

Furthermore, one can study tidal forces in the present of the two vector $Q^{\mu}$ and $Z^{\mu}$. If these vector vanish, all tidal effects are equivalent to the gravitational analogue, but in the non-vanishing case the vector should deform the dynamics creating a much richer situation.

We hope to verify all these possibilities in future works.

\section*{ACKNOWLEDGEMENTS}

One of us (F.T.Falciano) would like to thank CNPq of Brazil for financial support.  We would also like to thank `Pequeno Seminario' of CBPF's Cosmology Group for useful discussions, comments and  suggestions.\\

\appendix

\section{Generalisation for a non-inertial congruence in curved space-time}\label{appendixA}

We now consider the general case of a non-inertial observer in curved space-time. The covariant derivative of the vector field defining the observes can be characterised by the kinematical parameters of the congruence through
\[
V_{\mu ;\nu}=\frac{1}{3} \theta h_{\mu \nu}+\sigma_{\mu \nu}+w_{\mu \nu}+a_{\mu}V_{\nu}
\]
where $\theta \equiv V^{\lambda}_{\phantom a ; \lambda}$ is the isotropic expansion, $\sigma_{\mu \nu}$ is the shear, $w_{\mu \nu}$ is the torsion tensor and $a_{\alpha} \equiv V_{\alpha;\lambda}V^{\lambda}$ is the four-acceleration vector.

To find the divergence of $I_{\alpha \beta \lambda}$ in a Riemannian manifold, we recall the identity
\[
F^{\mu \nu ; \alpha \beta}=F^{\mu \nu ; \beta \alpha}+R_{\lambda}^{\phantom a \mu \alpha \beta}F^{\lambda \nu}+R_{\lambda}^{\phantom a \nu \alpha \beta}F^{\mu \lambda} \qquad .
\]

By contracting the $\nu$ and $\beta$ indexes and using the definition of $I^{\mu \beta \alpha}$ we find
\begin{equation}
I^{\mu \beta \alpha}_{\phantom a \phantom a \phantom a \, ;\beta}=\frac{2}{3}F^{\mu ; \alpha}+R^{\alpha}_{\phantom a \lambda}F^{\mu \lambda}+R_{\phantom a \beta \phantom a \lambda}^{\alpha \phantom a \mu }F^{ \beta \lambda} \qquad .
\end{equation}

The other dynamical equation can be found by taking the covariant derivative of the $I_{\alpha \beta \lambda}$ which combined with equations (\ref{max2}) and (\ref{defI}) give us
\begin{eqnarray*}
\nabla_{\left[ \alpha\right.}I_{\left. \mu \nu \right] \beta}&=&2R^{\lambda}_{\phantom a \beta \left[ \mu \nu \right.}F_{\left. \alpha\right] \lambda} -\frac{1}{3} \left(\nabla_{\left[ \alpha \right.}F_{\left. \mu \right]}g_{\nu \beta}\, +\right. \quad \\
&&\left.+\; \nabla_{\left[ \nu \right.}F_{\left. \alpha \right]}g_{\mu \beta}+\nabla_{\left[ \mu \right.}F_{\left. \nu \right]}g_{\alpha \beta}\right)\quad .
\end{eqnarray*}

Multiplying both sides of this equation by $\eta^{\rho \alpha \mu \nu}$ we find
\begin{equation}
I_{\phantom a \phantom a \phantom a ;\alpha}^{^{\phantom a \ast}_{\rho \alpha \beta}}=R^{^{\phantom a \phantom a \phantom a \ast}_{\lambda \beta \rho \alpha}}F_{\alpha \lambda}+\frac{1}{3}F^{^{\phantom a \ast}_{\rho ; \beta}}
\end{equation}

The dynamical equations related to the divergence of $I_{\alpha \beta \lambda}$ are generalised to 
\begin{widetext}
\begin{eqnarray*}
&\mbox{i)}I_{\phantom a \phantom a \phantom a ; \, \beta}^{\mu \beta \alpha}V_{\mu}V_{\alpha}:&-Q^{\beta}_{\phantom a \, ; \beta}-Q^{\mu}a_{\mu}+\left(E^{\beta \mu}+\frac{3}{2}\eta^{\mu \beta}_{\phantom a \phantom a \varepsilon \lambda}Z^{\varepsilon}V^{\lambda} \right)V_{\mu ;\beta}=\frac{2}{3}\left(\dot{\rho}-J^{\mu}a_{\mu} \right)\\
&\mbox{ii)}I_{\phantom a \phantom a \phantom a; \beta}^{\mu \beta \alpha}V_{\alpha}h^{\rho}_{\phantom a \mu}:&\dot{Q}^{\rho}+\theta Q^{\rho}-\left(Q^{\rho \mu}-V^{\rho}Q^{\mu} \right)a_{\mu}-\eta^{\rho}_{\phantom a  \varepsilon \sigma \beta}\left(Z^{\varepsilon}V^{\sigma} \right)^{; \beta}-V^{\rho}_{\phantom a ; \beta}Q^{\beta}-\left(V^{\rho}\eta^{\mu \beta}_{\phantom a \phantom a \varepsilon \sigma}Z^{\varepsilon}V^{\sigma}-\eta^{\rho \beta}_{\phantom a \phantom a \varepsilon \sigma}Z^{\varepsilon \mu}V^{\sigma} \right)V_{\mu ; \beta}=\\
&&=\frac{2}{3}\left[(j^{\rho}\dot{)}+J^{\mu}a_{\mu}V^{\rho}+\rho \, a^{\rho} \right]\\
&\mbox{iii)}I_{\phantom a \phantom a \phantom a; \beta}^{\mu \beta \alpha}V_{\mu}h^{\lambda}_{\phantom a \alpha}:&-E^{\lambda \beta}_{\phantom a \; ; \beta}-\left(Q^{\mu \lambda}-Q^{\mu}V^{\lambda}\right)a_{\mu}-\frac{1}{2}\eta^{\lambda}_{\phantom a \varepsilon \sigma \beta}\left(Z^{\varepsilon}V^{\sigma}\right)^{; \beta}-V^{\lambda}_{\phantom a \, ; \beta}Q^{\beta}+\\
&&-\left(E^{ \mu \beta}V^{\lambda}+\frac{3}{2}V^{\lambda}\eta^{ \mu \beta}_{\phantom a \phantom a \varepsilon \sigma}Z^{\varepsilon}V^{\sigma}-\eta^{ \mu \beta}_{\phantom a \phantom a \varepsilon \sigma}Z^{\varepsilon \lambda}V^{\sigma}\right)V_{\mu ; \beta}=\frac{2}{3}\left( \rho_{; \alpha}-J^{\mu}V_{\mu ; \alpha}\right)h^{\lambda \alpha}\\
&\mbox{iv)}I_{\phantom a \phantom a \phantom a; \beta}^{\mu \beta \alpha}h^{\rho}_{\phantom a \mu}h^{\lambda}_{\phantom a \alpha}:&\dot{E}^{\rho \lambda}+\theta E^{\rho \lambda}-\frac{1}{2}\eta^{\rho \lambda}_{\phantom a \phantom a \varepsilon \sigma}
\left( \dot{Z}^{\varepsilon}V^{\sigma}+Z^{\varepsilon}a^{\sigma}+\theta Z^{\varepsilon}V^{\sigma} \right)+Q^{\rho}a^{\lambda}+\left(E^{\beta \lambda}-\frac{1}{2}\eta^{\beta \lambda}_{\phantom a \phantom a \varepsilon \sigma}Z^{\varepsilon}V^{\sigma}\right)V^{\rho}_{\phantom a \, ; \beta}+\\
&&+\left( V^{(\rho}E^{\lambda)\mu}+\frac{1}{2}V^{(\rho}\eta^{\lambda) \mu}_{\phantom a \phantom a \phantom a \varepsilon \sigma}Z^{\varepsilon }V^{\sigma}\right)a_{\mu}-\eta^{\mu \beta}_{\phantom a \phantom a  \varepsilon \sigma}h^{\rho}_{\phantom a \mu}h^{\lambda}_{\phantom a \alpha}\left(Z^{\varepsilon \alpha}V^{\sigma}\right)_{; \beta}=\\
&&=\frac{2}{3}\left[\dot{\rho}V^{\lambda}V^{\rho}-\dot{J}^{\rho}V^{\lambda} -\rho^{\, ; \lambda}V^{\rho}+J^{\rho;\lambda}+J_{\mu}V^{\mu ;\lambda}V^{\rho}-J^{\mu}a_{\mu}V^{\lambda}V^{\rho}\right]\\
&\mbox{v)}I_{\phantom a \phantom a \phantom a ;\beta}^{^{\phantom a \ast}_{\mu \beta \alpha }}V_{\mu}V_{\alpha}:&-Z^{\beta}_{\phantom a ; \beta}-Z^{\mu}a_{\mu}+\left( H^{\mu \beta }-\frac{1}{2}\eta^{\mu \beta}_{\phantom a  \phantom a  \varepsilon \lambda}Q^{\varepsilon}V^{\lambda}\right)V_{\mu ; \beta}=0\\
&\mbox{vi)}I_{\phantom a \phantom a \phantom a \, ;\beta}^{^{\phantom a \ast}_{\mu \beta \alpha}}V_{\alpha}h^{\rho}_{\phantom a \mu}:&\dot{Z}^{\rho}+\theta Z^{\rho}-\left(Z^{\rho \mu}-Z^{\mu}V^{\rho}\right)a_{\mu}+\eta^{\rho}_{\phantom a  \mu \nu \beta}\left(Q^{\mu}V^{\nu}\right)^{; \beta}-V^{\rho}_{\phantom a \, ; \beta}Z^{\beta}+\left(V^{\rho}\eta^{\mu \beta}_{\phantom a \phantom a \varepsilon \sigma}Q^{\varepsilon}V^{\sigma}-\eta^{\rho \beta}_{\phantom a  \phantom a  \varepsilon \sigma}Q^{\varepsilon \mu}V^{\sigma} \right)V_{\mu ; \beta}=\\
&&=\frac{1}{3}\eta^{\rho}_{\phantom a \varepsilon \sigma \alpha}J^{\varepsilon \, ; \sigma}V^{\alpha}\\
&\mbox{vii)}I_{\phantom a \phantom a \phantom a ;\beta}^{^{\phantom a \ast}_{ \mu \beta \alpha}}V_{\mu}h^{\lambda}_{\phantom a \alpha}:&-H^{\lambda \beta}_{\phantom a \phantom a ;\beta}-\left(Z^{ \mu \lambda}-Z^{\mu}V^{\lambda}\right)a_{\mu}-\frac{1}{2}\eta^{\lambda}_{\phantom a  \varepsilon \sigma \beta}\left(Q^{\varepsilon}V^{\sigma}\right)^{ ; \beta}-V^{\lambda}_{\phantom a \, ; \beta}Z^{\beta}+\\
&&-\left(H^{\mu \beta}V^{\lambda}-\frac{1}{2}V^{\lambda}\eta^{\mu \beta}_{\phantom a \phantom a \varepsilon \sigma}Q^{\varepsilon}V^{\sigma}-\eta^{\mu \beta}_{\phantom a  \phantom a \varepsilon \sigma}Q^{\varepsilon \lambda}V^{\sigma} \right)V_{\mu ; \beta}=-\frac{1}{3}\eta^{\lambda}_{\phantom a \varepsilon \sigma \alpha}J^{\varepsilon \, ; \sigma}V^{\alpha}\\
&\mbox{viii)}I_{\phantom a \phantom a \phantom a ;\beta}^{^{\phantom a \ast}_{\mu \beta \alpha}}h^{\rho}_{ \phantom a \mu}h^{\lambda}_{\phantom a \alpha}:&\dot{H}^{\rho \lambda}+\theta H^{ \rho \lambda}-\frac{1}{2}\eta^{\rho \lambda}_{\phantom a \phantom a \varepsilon \sigma}\left(\dot{Q}^{\varepsilon}V^{\sigma}+Q^{\varepsilon}a^{\sigma}+\theta Q^{\varepsilon}V^{\sigma}\right)+Z^{\rho}a^{\lambda}-\left(
H^{\beta \lambda}-\frac{1}{2}\eta^{\beta \lambda}_{\phantom a \phantom a \varepsilon \sigma}Q^{\varepsilon}V^{\sigma}\right)V^{\rho}_{\phantom a \, ; \beta}+\\
&&+\left(V^{(\rho}H^{\lambda)\mu}+\frac{1}{2}V^{(\rho}\eta^{\lambda) \mu}_{\phantom a \phantom a \; \varepsilon \sigma}Q^{\varepsilon}V^{\sigma}\right)a_{\mu}+\eta^{\mu \beta}_{\phantom a \phantom a  \varepsilon \sigma}h^{\rho}_{\phantom a \mu}h^{\lambda}_{\phantom a \alpha}\left(Q^{\varepsilon \alpha}V^{\sigma}\right)_{; \beta}=\frac{1}{3}\left(\eta^{\rho \lambda}_{\phantom a \phantom a \varepsilon \sigma}J^{\varepsilon \, ; \sigma}+V^{\left[ \rho \right.}\eta^{\left. \lambda \right]}_{\phantom a \varepsilon \sigma \alpha}J^{\varepsilon \, ; \sigma}V^{\alpha} \right)
\end{eqnarray*}
\end{widetext}



\begin{thebibliography}{99}

\bibitem{Ruffini} Remo J. Ruffini ``Introduction to non-linear gravitodynamics: the Lense-Thirring effect'', in Non-linear gravitodynamics: The Lense-Thirring effect (a documentary introduction to current research), world scientific, edited by Remo Ruffini and Constantino Sigismondi (2003).
\bibitem{deSitter} W. de Sitter, Mon. Not. Roy. Astron. Soc. 76, 699 (1916).
\bibitem{Mashoon} Bahram Mashoon, Friedrich W. Hehl, and Dietmar S. Theiss, ``On the gravitational effects of rotating masses: The Thirring-Lense papers'' Gen. Rel. and Grav. vol 16, 8, (1984); H. Thirring, Phys. Z. 19, 33 (1918); 22, 29 (1921); J. Lense and H. Thirring, Phys. Z., 19, 156 (1918);
\bibitem{Sciama}, D. W. Sciama, Mon. Not. Roy. Astr. Soc. vol. 113, pp. 34-42, (1953).
\bibitem{Pugh} G. E. Pugh ``Proposal for a Satellite Test of the Coriolis Prediction of General Relativity'', WSEG Research Memorandum no 11 (1959).
\bibitem{exemplos} Edward G. Harris; ``Analogy between general relativity and electromagnetism for slowly moving particles in weak gravitational fields'' Am. J. Phys. 59 (5), May (1991); M. Novello et al, ``Electric and magnetic gravitational monopoles I. The equation of motion of poles'' J. Phys. A: Math Gen.,Vol. 9, No 4, 547 (1976); Forward, Robert L., Proc. IRE 49, 892 (1961); R. Maartens and B.A. Bassett, Class. Quantum Grav. 15, 705 (1998); S. J. Clark and R. W. Tucker, Class. Quantum Grav. 17, 4125 (2000); P. Teyssandier, Phys. Rev. D 16, 946 (1977); B. Mashhoon, Phys. Lett. A 173, 347 (1993) / Phys. Lett. A 139, 103 (1989); S. Kopeikin and B. Mashhoon Phys. Rev. D 65, 064025 (2002). B. Mashhoon, Class. Quantum Grav. 25, 085014 (2008).
\bibitem{testes} Vladimir B. Braginsky, Carlton M. Caves and Kip S. Thorne; ``Laboratory experiments to test relativistic gravity'' Phys. Rev. D volume 15, number 8, 2051 (1977); I. Ciufolini, ``Dragging of inertial frames'' Nature Vol 449, 41 (2007); S. M. Kopeikin, E.B. Fomalont, Gen. Rel. Grav. 39, 1583 (2007); B. Mashhoon, Physics Letters A 198, 9 (1995); H. I. M. Lichtenegger, F. Gronwald and B. Mashhoon,  Adv. Space Research 25, 1255 (2000); C. A. Blockley and G. E. Stedman, Physics Letters A 147, 161 (1990); A. Camacho, (2002), Gen. Rel. Grav., 34, (1963); I. Ciufolini, Phys. Rev. Lett., 56, 278 (1986); I. Ciufolini and E. C. Pavlis, Nature, 31, 958 (2004) / New Astron., 10, 636 (2005); N. Ashby, Nature, 31, 918 (2004); L. Iorio, New Astron., 10, 603 (2005); C. M. Will Phys. Rev. D 67, 062003 (2003); S.M. Kopeikin, Int. J. Mod. Phys. D, 15, 305 (2006).
\bibitem{Wein-Ohan} Weinberg, S., ``Gravitation and Cosmology'', John Wiley and Sons, New York, (1972); Hans C. Ohanian, Remo Ruffini, ``Gravitation and Spacetime'', W.W. Norton \& Company, second edition (1994).
\bibitem{Lichnerowicz} Lichnerowicz A.  Ann. Mat. Pura Appl. 50, 1 (1960).
\bibitem{Ellis} G. F. R. Ellis, General Relativity and Cosmology, Proc. of the Int. School of phys., Enrico Fermi XLVII (Academic, London), p. 104 (1971).
\bibitem{Nov2} M. Novello and J. M. Salim, ``Non-equilibrium relativistic cosmology'' Fundam. of Cosm. Phys. 1983, Vol 8, pp. 201-342.
\bibitem{Bel} Louis bel, General Relativity and Gravitation, vol. 32, 10 2000. 
\bibitem{Hawk} S. W. Hawking, Ap Journal 145, 544. 1966.
\bibitem{Choquet} Y. Choquet-Bruhat, C. Dewitt-Morette, M. Dillard-Bleick ``Analysis, Manifolds and Physics'', Elsevier Science Pub Co (Revised edition) (1996).
\bibitem{Matte} A. Matte Can. J. Math. Vol. 5, p.1 (1953).
\bibitem{Jordan} P. Jordan, J. Ehlers and W. Kundt  Abh. Math.-Naturw. Kl. Akad. Wiss. Mainz No2 (1960).
\bibitem{Kundt} W. Kundt, M. Trumper, Abh A Mat. Nat. k1 12 (1960)
\bibitem{Salim} M. Novello et al ``Electric and magnetic monopoles'' J. Phys. A: Math. Gen. Vol 9, No4, (1976).
\bibitem{Novello4} M. Novello ``Lanczos Potential and Jordan theory of gravity'' , CBPF physics notes NF-054 (1983).
\bibitem{Nov3} M. Novello, J. Duarte de Oliveira, ``On dual properties of the Weyl tensor'' Gen. Rel. and Grav., Vol. 12, No.11, (1980).
\bibitem{Gravitodynam} Non-linear gravitodynamics: The Lense-Thirring effect (a documentary introduction to current research), world scientific, edited by Remo Ruffini and Constantino Sigismondi (2003).
\bibitem{Teves} J. D. Bekenstein, Phys. Rev. D 70, 083509 (2004); J. D. Bekenstein and  R. H. Sanders arXiv:astro-ph/0509519v1.
\bibitem{Deb} R. Debever, Cah. Phys. 168, 303 (1964)
\bibitem{Papapetrou1}  A. Papapetrou, ``Spinning test-particles in general relativity I''. Proc. Roy. Soc. London A 209, 248-258 (1951)
\bibitem{Papapetrou2} E. Corinaldesi and A. Papapetrou ``Spinning test-particles in general relativity II''. Proc. Roy. Soc. London A 209, 259-268 (1951).
\bibitem{falcianogoulart} F.T. Falciano and E. Goulart, in preparation. 
\bibitem{costaherdeiro} L. Filipe Costa and Carlos A. R. Herdeiro; ``A gravito-electromagnetic analogy based on tidal tensors'' arXiv:gr-qc/0612140v2 21 May (2007).
\end{thebibliography}
\end{document}